\input epsf.sty

\def\beq{\begin{equation}}
\def\eeq{\end{equation}}
\def\bea{\begin{eqnarray}}
\def\eea{\end{eqnarray}}
\def\bq{\begin{quote}}
\def\eq{\end{quote}}

\def\gappeq{\mathrel{\rlap {\raise.5ex\hbox{$>$}}
{\lower.5ex\hbox{$\sim$}}}}

\def\lappeq{\mathrel{\rlap{\raise.5ex\hbox{$<$}}
{\lower.5ex\hbox{$\sim$}}}}

\def\bbz{fa Z \kern-8.9pt Z}

\documentstyle [12pt]{article}
\begin{document}
\thispagestyle{empty}
\vspace*{-2cm}
\begin{flushright}
{CERN-TH/98-324} \\
{October 1998} \\
\end{flushright}
\vspace{1cm}
\begin{center}
{\large Baryogenesis in Models with a Low Quantum Gravity Scale} \\
\vspace{.2cm}
\end{center}
\vspace{1cm}
\begin{center}
{ Karim Benakli and Sacha Davidson }\\
\vspace{.3cm}
{CERN Theory Division\\
 CH-1211, Gen\`eve 23, Switzerland}
\end{center}
\hspace{3in}

\begin{abstract}
We make generic remarks
about baryogenesis in models where the scale $M_s$
of quantum gravity
is much below the Planck scale. These correspond to M-theory vacua
with a large volume for the internal space. 
Baryogenesis is a challenge,
particularly for $M_s \lappeq 10^5$ GeV,
because there is an  upper bound 
on  the reheat temperature of the Universe, and  
because certain baryon number violating operators 
  must be suppressed.
We discuss these constraints  for different values of $M_s$,
and  illustrate with a toy model the possibility of using 
horizontal family symmetries to circumvent them.
\end{abstract}

\section{Introduction}

There are three experimental observations that might  be considered as
evidence
for beyond-the-Standard-Model physics: neutrino oscillations,
 \cite{nu98}
the Baryon Asymmetry of the Universe (BAU),
\cite{BAUrev}
and the temperature fluctuations
in the microwave background 
\cite{CMB}.
Any extension of the Standard Model must
explain, or at least be consistent with, this data.

One of the reasons to attempt to extend the Standard Model is the
possibility of unifying gravity with the other interactions.  Present
candidates are believed to be vacua of a single fundamental  theory:
M--theory.  The formulation of the latter  seems to require adding
new degrees of freedom. In a regime where  a semi-classical
description holds, these degrees of freedom  manifest  themselves  as
additional spatial dimensions compactified into an internal space.
In its present form, M--theory makes no prediction about the size of 
any spatial dimensions. It allows certain vacua with arbitrary 
large size for the internal dimensions
limited only by experimental data. If the  states
propagating in these dimensions have  couplings with size  comparable
to those of standard model gauge interactions then  the
non-observation of effects  associated with Kaluza-Klein  excitations
leads to lower limits on the size of internal radii of the order of 
$\sim$ TeV \cite{TeVlim}. If, in contrast, all the
couplings of these Kaluza-Klein  excitations are of the strength of
gravitational interactions, then the limit is of around a
millimetre \cite{mmlim} \footnote{ Notice that the scale suppressing the
interactions has increased by 15 orders of magnitude and the 
experimental limits went down with roughly the same amount.}. Mechanisms
for stabilization of the radii of the extra-dimensions have been discussed 
in \cite{ADMR}.

Allowing the presence of such large internal dimensions has  dramatic
effects on  phenomenological aspects of M--theory. Above the scale
where the  largest dimensions  lie, naive dimensional analysis shows
that the strength of gravitational  interactions increases rapidly
with  energies. This implies that gravity and the  three other known
fundamental interactions will have the same strength and might unify
at a scale $M_s$ than can be very low TeV $\lappeq M_s\lappeq 10^{19}$
GeV.  At $M_s$ quantum gravity effects become important and new
unknown phenomena might arise.  Remnants of these phenomena at low
energies are various non-renormalisable effective operators. The size
of the latter,  if observed, might provide an indication 
on the existence
and range  of values of $M_s$.

This possibility of a low quantum gravity scale  was first suggested 
in \cite{witten}
with a scale $M_s$  at $\sim  10^{16}$ GeV leading to  unification
within the Minimal Supersymmetric Standard Model of all the
interactions. It was later observed that  Type I strings
\cite{typeI} (also motivated also by a field theoretical 
proposal in \cite{ADD0} and for which model building was studied in \cite{tye}),
 M--theory on $S^1/Z_2$ \cite{kar} and possibly 
heterotic strings \cite{bachas0} allow $M_s \sim $ TeV. 
  This opens the exciting possibility that
extra-dimensions could  be observed at future colliders \cite{ABK}. Another 
proposal is to
have $M_s$ at an intermediary scale \cite{kar}  so
as to be associated with neutrino
masses, observed ultra-high energy cosmic rays or the scale of breaking of a
Peccei-Quinn symmetry. In this
case the standard unification scenario might also be preserved
\cite{bachas}.

In addition to the early phenomenological bounds for large internal dimensions 
 discussed above, 
other limits on $M_s$ have recently been derived \cite{ADD} from
astrophysical and cosmological considerations. The most
significant particle physics constraint on $M_s$ 
that we are aware of  comes from atomic parity 
violation experiments \cite{APV}, which determine
$\sin^2 \theta_W$ at low energy. 
If we assume\footnote{Note
that we do not use the common $2 \pi/\Lambda^2$
normalisation of the new physics contribution
to the four fermion vertex. Had we done so, we would
have found $M_s > 10-14$ TeV.} that  the coefficient of
the four fermion vertex $4G_F/\sqrt{2}$ 
becomes $4G_F/\sqrt{2} + 1/M_s^2$, 
we get $M_s > 4-6$ TeV. The strongest
astrophysical bound estimated in \cite{ADD} is from
supernovae, and requires
$M_s \gappeq  30$ TeV in the case of two large 
compactified dimensions.

The purpose of this paper is to investigate consequences of these models for
baryogenesis. We will restrict our study to the class of models
where matter and gauge fields live on a 3+1 dimensional 
wall and interact only through weak interactions 
of gravitational strength with fields living
in the $(3+n)$+1 dimensional ``bulk''. The thermodynamics for 
the case with gauge interactions in higher dimensions (bulk) was recently 
studied in \cite{voisins}. In the absence 
of a precise model, we introduce three
mass parameters in various stages of our analysis. 
The first is $M_s$ where gravity unifies 
with the other interactions. 
It corresponds to the string scale in string models 
or to the eleven-dimensional 
Planck mass in Ho\v rava-Witten \cite{HW} 
compactification of M--theory. 
The second is $m_{pl(4+n)}$ which is the Planck scale 
in $(3+n)+1$ dimensions. The relation 
between $m_{pl(4+n)}$ and $M_s$  involves
the volume of the  dimensions with 
smaller radii. If the latter are of order $M_s^{-1}$
then $m_{pl(4+n)} \sim M_s$. 
Another parameter that we generically denote by $\Lambda$
appears as a suppression scale for 
different non-renormalisable operators. It is related to 
$M_s$ through model dependent coupling constants and numerical factors.

In section two we discuss experimental
bounds on non-renormalisable baryon number violating
operators, and which operators
need to be forbidden for different
values of $\Lambda$. In section three
we make some remarks about inflation,
and discuss the upper bound on the
reheat temperature of the Universe  $T_{reh}\ll M_s$
that follows from the production of
gravitons in the large internal dimensions.
 Graviton production
during the reheating period is dangerous as their
decay products  can lead to  a greater
than observed  differential photon flux. 
In section 4 we discuss  the difficulties 
of reconciling baryogenesis with the  
 suppression of baryon 
number violating operators and the
upper bound on the reheat temperature.
We consider the possibility 
of generating the baryon asymmetry in
the  out-of-equilibrium 
decay of  a weakly coupled particle.
To provide sizable decay channels we suggest using
horizontal family symmetries to suppress dangerous non-renormalisable 
operators instead of forbidding them through (discrete) gauge symmetries.
A toy model for baryogenesis is exhibited to illustrate this scenario.
Section 5 summarises our conclusions.

\section{Baryon number violating operators}

The presence of new physics at low scales
 could generate dangerous non-renormalisable
operators. These could for instance  lead to unobserved
baryon number violating processes
such as proton decay and neutron-anti-neutron
oscillations. In the absence of a precise
model, where such operators can be computed,
we make the conservative assumption that every 
 operator 
that is not forbidden by a  (possibly discrete) 
gauge symmetry
could be  generated  with a coefficient
of order one\footnote{In this work, we 
apply this assumption to $B$ and $L$
violating operators, but not, for instance, to
FCNC.}. 
This means that non-renormalisable  
baryon number violating operators
of dimension $4+d$ could  appear, suppressed
by factors of 
 the scale of new physics $M_s$. The
precise coefficient  of a $4+d$ dimensional
operator will involve  $M_s$, various
coupling constants and numerical factors, which
we absorb into a coefficient called  $\Lambda^{-d} $.

A strong constraint on baryon
number violating operators
  is that the proton
must have a lifetime $\tau_p \gappeq 10^{33}$ years
\cite{SUPERK}.
If the quantum gravity
scale is low, this means that
one must forbid
baryon and lepton number violating operators
up to some large dimension \cite{mmlim,proton}. For instance the
operator $(QQQL)/\Lambda $ in supersymmetry (SUSY)
generates proton decay at a rate of
order \cite{GR}
\beq
\Gamma \sim 10^{-2} \frac{\alpha^2 m_p^5}{\Lambda^2 m_{SUSY}^2}
\eeq
which implies $\Lambda \gappeq 10^{26}$ GeV (!).
For non-supersymmetric models,
 the operator is dimension 6 and the
bound becomes  $\Lambda \gappeq 10^{15}$ GeV.

Another  baryon number violating
process that presents a significant 
constraint for low $M_s$ is
neutron-anti-neutron oscillations.
This is a $\Delta B = 2$, $\Delta L = 0$
process, that is generated by the
dimension 9 operator $udsuds$.
The ``lifetime'' for neutron-anti-neutron
oscillations $\tau_{n\bar{n}} > 1.2 \times 10^{8}$ seconds
\cite{PDG} is of order
\beq
\tau_{n\bar{n}} \simeq \frac{\Lambda^5}{5 \times 10^{-6} {\rm GeV}^6}
\eeq
in the SM,
where the denominator is an estimate of the
hadronic matrix element \cite{Zwirner,had}.
This gives $\Lambda \gappeq 10^{5}$ GeV.

A   list of baryon and lepton
number violating operators  in the Standard
Model (SM) and the Minimal Supersymmetric Standard Model
(MSSM) is given in table 1
with  approximate bounds on the scale $\Lambda$.
One must forbid 
with some symmetry all operators that are experimentally
constrained to have $\Lambda > M_s$.

We follow \cite{GR} 
to calculate the constraints in the table.
We take all supersymmetric particle masses
and Higgs vevs to be 100 GeV, and the hadronic
matrix elements for proton  decay 
to be $\sim 10^{-2}$ (with appropriate
mass dimensions provided by the proton mass).
The table is not particularly illuminating, because the bounds
do not simply scale with the dimension. Roughly, operators
that violate $B$  and L by one unit each are forbidden
up to scales $>10^{10}$ GeV, operators that violate
$B$ alone by one or two units are forbidden up to scales
of order $10^{5}$ ($10^{9}$) GeV in the SM (MSSM), 
and operators that violate
$B$  by three units are allowed at the TeV scale. 
An example of a symmetry  that forbids
$\Delta B = 1$ and 2 processes in the MSSM  is 
the discrete anomaly-free $Z_3$ symmetry
of Iba\~nez and Ross \cite{IR} which conserves $B$ mod 3.
 The lowest
baryon number violating operators it allows 
are combinations like $(QQQL)^3, (U^cU^cD^cE^c)^3$ and
$(QQQH_1)^3$.

\begin{table}
\begin{tabular}{||c|c|c|c|c|c|c||} 
\hline
\hline
{\rm operator} & &{\rm process} & {\rm SUSY~dim} & {\rm SUSY~bd} 
   & {\rm SM~dim}  & {\rm SM~bd} \\
\hline
$Q_1Q_1Q_2L$ & $\Delta B = \Delta L = 1$&$ p \rightarrow K \nu$ 
  & 5 & $  10^{26} $
  & 6 &$  10^{15}$\\ 
  \hline
$U_1^cU_2^cD_1^cE^c$ &$\Delta B = \Delta L = 1$&$ p \rightarrow K \nu$ 
  & 5  & $ 10^{22}$ 
  & 6  &  $  10^{12}$ \\ 
  \hline
$Q_1Q_1Q_2H_1$ &$\Delta B = 1$& $ n - \bar{n} $ 
  & 5 & $ 10^{9}$
  & --- & ---\\
  \hline
 \hline 
$U_1^cU_2^cU_3^cE^cE^c$ &$ \Delta B = 1,\Delta L = 2$&? 
  &6  & 
  &--- &--- \\
  \hline
$U_1^cD_1^cD_2^cH_1H_2$ &$\Delta B = 1$& $ n - \bar{n} $  
  &6  & $10^{5}$
  &--- &--- \\
  \hline
$D_1^cD_2^cD_3^c L H_1$ &$\Delta B = - \Delta L = 1$&$ n\rightarrow \nu \pi$
  &6  & $ 10^{13}$ 
  &7  & $10^{9}$ \\
  \hline
$U_1^cD_1^cD_2^c L H_2$ &$\Delta B = - \Delta L = 1$
  &$ n\rightarrow \nu K$ 
  &6  & $ 10^{14}$ 
  &7  & $ 10^{10}$ \\
  \hline
  \hline
$U_1^cD_1^cD_2^cU_1^cD_1^cD_2^c$ &$ \Delta B = 2$& $ n - \bar{n} $  
  &7  & $10^5$ 
  &9  &$ 10^{5}$ \\
 \hline
$U_1^cD_1^cD_2^cLLE^c $&$\Delta B = - \Delta L = 1$& 
  $n \rightarrow e^+ \mu^- \nu$ 
 & 7 & $6\times 10^{7}$
 &9  & $5 \times 10^{5}$\\
 \hline
$U_1^cD_1^cD_2^cLQD^c $ &$ \Delta B = - \Delta L= 1$& 
  $ n \rightarrow e^+ \pi  $
   &7  & $10^7$
   &9 & $4 \times 10^{5}$ \\
 \hline
$U_1^cU_2^cD_1^cH_2LE^c $ &$ \Delta B = 1$& $ n - \bar{n} $
   &7  & $\lappeq 10^{3} $
   &--- &---  \\
 \hline
$U_1^cU_2^cD_1^cH_2QD^c$ &$ \Delta B = 1$& $ n - \bar{n} $  
  & 7 &  $ \lappeq 10^{3} $
  &--- &--- \\ 
 \hline
$QQQLLH_2$ &$\Delta B = 1,\Delta L = 2$& 
 ? & 7 &  ?
  &--- &--- \\ 
 \hline
 \hline
$Q_1Q_1Q_2H_1Q_1Q_1Q_2H_1$ & $\Delta B = 2$ &$n - \bar{n} $& 9 
  & $ 10^{4}$  
  & 11 & $10^4$ \\ 
 \hline
 \hline
\end{tabular}
\label{t1}
\caption{B  violating operators of dimension $> 4$
for Standard Model and MSSM particle content, in 
superfield notation. These are only the 
``F-terms''. We list the dimension of the
operators, the processes they contribute to,
and the best bound we are aware of (in GeV), assuming that
the coefficient of a dimension $d+4$ operator is $\Lambda^{-d}$.
The quark  field subscripts are
generation indices.
We do not include operators of the form
(allowed lower dimensional operator) $\times$ (forbidden
lower dimensional operators), such as $LH_2H_1H_2$
or $U^cD^cD^cLH_1E^c$, because they are forbidden
be whatever removes the unwanted lower dimensional
operator. }
\end{table}

Note that the bounds on the operators
in table 1
are usually for first generation  quarks and
leptons. For low quantum gravity scales,  some sort of flavour
symmetry presumably should be imposed to remove FCNC
operators, so one could imagine that
that there are flavour dependent symmetries
that forbid or suppress the dangerous 
baryon and/or lepton number
violating operators. For instance,
if the hierarchy in the
yukawa couplings is due to a
spontaneously broken horizontal
symmetry \cite{hor}, the baryons and leptons 
can be assigned charges under
this symmetry  that 
forbid most of the operators in table 1
({\it e.g.} by giving all the SM fermions positive
charges).  We will discuss this possibility in section 4.2.

\section{Inflation and reheating}

\subsection{Inflation}

A period of inflation is the only known way of generating 
the temperature fluctuations measured
on scales up to 100 Mpc in the microwave background.
Since inflation dilutes any pre-existing  asymmetries,
the observed Baryon Asymmetry of the Universe
(BAU) must be generated afterwards. 
As we will see, there is an upper bound 
on the  reheat temperature
in models with low quantum gravity scale,
 so the phase transition
out of inflation is one of the few
places where one can find the out-of-equilibrium
required for baryogenesis.

If we take the energy density of the Universe to be
at most $M_s^4$, then  for $n \geq 2$
 large internal dimensions,
the Hubble radius is greater
than or equal to the  radius of
the $n$ dimensions.
This means that it is consistent
to build an inflation  model in
 3+1 dimensions. However, a second order inflation model
at a scale $\ll 10^{15}$ GeV requires a great
deal of fine tuning to get enough
e-foldings and the density perturbations of order
$10^{-5}$.
The latter can be estimated as
\beq
\frac{\delta \rho}{\rho} \sim \frac{V^{3/2}}{m_{pl}^3 V'}
\eeq
where $V$ is the potential energy density
of the inflaton,  $V' = dV/d \phi$, and both
of these are evaluated at the point in the potential
where the inflaton was sitting 50 - 60 e-folds before
the end of inflation.  If $V \sim M_s^4$,
then 
\beq
\frac{V'}{M_s^3} \sim 10^{5} \left(\frac{M_s}{m_{pl}} \right)^3
\eeq
so the potential must be very, very flat.
If, for instance, one parametrises $V = V_0 - m^2 |\phi|^2 +
\lambda |\phi|^4 +\sum \phi^{n+4}/M_s^n$,  with
$V_0 \sim M_s^4$, then
to get enough inflation \cite{T,RL} and the right sized density
perturbations, one finds $m \sim M_s^2/m_{pl}$.
 For $M_s \sim$ TeV,
one gets $m \sim 10^{-13}$ GeV.  Such
a light inflaton might have difficulties
reheating the Universe to temperatures $\sim$ MeV,
and in any case, $V_0 \sim m^4 \ll M_s^4$,
so our initial assumptions were inconsistent.
 To avoid this difficulty, one can
build two field or hybrid inflation models \cite{RL}
where the mass of the inflaton when it decays
is not related to the mass term in the potential
when it is generating density perturbations.
An ad hoc  potential of the form 
$V_0 - a_6 \phi^6/M_s^2 + a_{12} \phi^{12}/M_s^8$
also works, for $a_6 \sim a_{12} \sim 10^{-2}$ and $M_s \sim 10$ TeV.
For the rest of this work, we will assume that  the potential 
is flat enough to inflate
for long enough, and that the mass of the inflaton
when it decays might be greater than a GeV.
This is useful for baryogenesis, if we
want to generate the asymmetry in the decay
of the inflaton.

\subsection{Gravitons production constraints on $T_{reh}$}

The Universe must at some point get out
of its inflationary phase, and reheat
to a plasma of particles. A safe
reheat temperature $ T_{reh}$ to ensure that
primordial nucleosynthesis
takes place as usual is $\gappeq$ 3 MeV \cite{lowT}. 
Baryogenesis at such a low
energy scale is hard, so a
higher $T_{reh}$ would be desirable.

Getting a high $T_{reh}$ is a challenge
in low  quantum
gravity scale models where the matter
lives on a 3+1 dimensional ``wall'',
while gravitons and other very
weakly interacting particles live
in the (3+$n$)+1 dimensional ``bulk''.  
The temperature to which the
Universe reheats must be low, to avoid
generating too many ``bulk particles''
(we will generically refer to them as
gravitons) in the extra
large dimensions. These gravitons
can decay  into  particles in
our 3+1 dimensional boundary. We can set bounds
on the number of these decay products
from various observations, and therefore
set an upper bound on the number of gravitons
allowed, or equivalently, an upper bound
on the reheat temperature $T_{reh}$. Below,
we estimate this bound as a function of the quantum
gravity scale and the number of large extra dimensions.

The behaviour of gravitons when $M_s
\sim$ TeV was discussed in
 \cite{ADD}. Their best bound comes from
requiring that photons from graviton
decay do not generate a spike  in the
$E \gg 2.7 ~^oK$ photon background.
For larger $M_s$, fewer gravitons are produced
so higher reheat temperatures are allowed. However, as the
graviton lifetime  becomes shorter, 
the decay products arrive in our 3+1 dimensions
at earlier epochs, so the limit on their
number density changes. If the gravitons
decay between recombination and today,
the photons produced will be in the
present photon background. For some period before
recombination, photon number changing
interactions in the thermal
plasma are out of equilibrium,
so photons from graviton decay produced
at this time would generate a chemical potential
for the microwave background. If the gravitons
 decay before recombination but after nucleosynethesis,
they can dissociate light elements. The bound
from this is similar to  the one from the
chemical potential.  Gravitons that decay before
nucleosynthesis are not a problem.
 We discuss bounds for all cases below.

One assumption made is that translation invariance in the bulk 
is broken only at the boundaries. This allows us to speak about momenta 
and energy of particles living in the bulk. 
Such a situation is not generic 
as the size of other dimensions 
of the internal space might become larger 
when going away from our wall towards a hidden one 
(see for instance \cite{kar}). It was argued in 
\cite{ADD} that gravitons might decay 
earlier on the hidden wall than on 
the observable wall avoiding some of 
our constraints. We will discuss this situation 
elsewhere \cite{BD2}.

Consider first the number density 
$n_G$ of gravitons produced in the bulk.
We follow \cite{ADD}
(a similar analysis was done in \cite{Steve}), 
and assume that the cross-section
for particles on the wall
to produce gravitons in the $n$ extra large
dimensions is of order\footnote{ It is the  
$4+n$--dimensional
``Planck scale'' $m_{pl (n+4)}$ that  appears, if
we assume that the other internal
dimensions have size of the order of $M_s^{-1}$ then
$m_{pl (n+4)}\sim M_s$.} 
$\sigma_{\gamma \gamma \rightarrow GG} \sim {T^n}/m_{pl (n+4)}^{n+2}$,
so the rate at which gravitons are made is approximately
\beq
\frac{\partial n_G}{\partial t} - 3 H n_G = 
\sigma n_{\gamma} \sim \frac{T^{n+6}}
{m_{pl (n+4)}^{n+2}}~~~.
\eeq
where $H$ is the Hubble expansion rate $H^2=8\pi \rho / 3m_{pl}^2$ and 
$n_{\gamma}$ is the number density of photons.
Gravitons made at a temperature $T$ will 
have  momenta in the bulk of order $T$, and since these momenta
do not redshift, the energy of the gravitons
remains $\sim T$. The number density of gravitons with
energy $T$ at later times (when the photon temperature
is $T_{\gamma}$) will therefore
be of order 
\beq
n_G(E=T) \simeq \sigma n_{\gamma} H^{-1}(T)  = N \frac{ m_{pl} T^{n+1} }
{m_{pl (n+4)}^{n+2} } T_{\gamma}^3 
\eeq
where $N$ is a numerical coefficient which we
have not calculated, and $m_{pl}$ is the 3+1
dimensional Planck mass. We take $N= 1$ in
figure 1. The number and energy of the gravitons
increases with $T$, so the most troublesome ones
are those generated at the reheat temperature
$T_{reh}$. We concentrate on these and consider constraints 
for different values of $m_{pl (n+4)}$.

For the lowest values  of $M_s$,
the strongest constraint obtained  in \cite{ADD}
on the number density of gravitons
is from the decay of gravitons back into
photons. We review this bound here. The
gravitons of energy $E$ 
decay to photons of energy $\sim E$
at a rate \cite{BEN}
\beq
\Gamma_G = \tau_G^{-1} = D \frac{E^3}{m_{pl}^2}
\eeq
where $D$ is another unknown numerical factor
that we set to 1 in figure 1.
 For $E \sim T_{reh} \lappeq 60 D^{-1/3} $ MeV,
 the lifetime of the graviton 
$\tau_G$ is longer than the age of the Universe $ \tau_U$.
The number who will have decayed is therefore
of order $n_{G0}\tau_{U}/\tau_G$.
 Following \cite{K+T},  one
 can require that the
flux of photons of energy $T_{reh}$ from
these decays not exceed the 
observed differential photon flux ${\cal F}$:
\beq
\frac{n_{G0}}{4 \pi} \frac{\tau_{U}}{\tau_G} \lappeq {\cal F}(E) = 
\frac{\rm MeV}{ E} {\rm cm}^{-2} {\rm sr}^{-1} {\rm sec}^{-1}
\eeq
where $E$ is the photon energy.
This gives
\beq
\frac{ND}{6 \pi} \frac{T_0^3}{H_0 m_{pl}} \left( \frac{T_{reh}}{m_{pl (n+4)}}
\right)^{n+2} (T_{reh})^2 < {\cal F}(T_{reh})
\eeq
where $T_0$ is the microwave background temperature
today.
This implies  
\beq
(T_{reh})^{n+5} <\frac{7 \times  10^{-39}}{ND}
m_{pl (n+4)}^{n+2} GeV^3 ~~~~~({\rm for~} T_{reh} < 60
  ~{\rm MeV})
\eeq
For $n=2$ and $T_{reh} \gappeq  3$  MeV 
(a safe reheat temperature to
ensure that primordial nucleosynthesis 
takes place as usual \cite{lowT}),
we get $m_{pl (n+4)} > 100 $ TeV.

For $T_{reh} > 60D^{-1/3}$ MeV, the gravitons
created at $T_{reh}$ can decay before
today. All their energy is therefore
in the photon background, but redshifted
from when they decayed until now.
If this took place after recombination,
we can set a bound by requiring
that their final  products do
not exceed the observed photon
flux ${\cal F}$.
The photon temperature $T_d$ when
the gravitons decay  can be computed from
\beq
H(T_d) \simeq \frac{2 T_{eq}^{1/2} T_d^{3/2}}{m_{pl}} \simeq \Gamma_G \simeq 
 D \frac{T_{reh}^3}{m_{pl}^2}
\eeq
where $T_{eq} \sim 3$ eV is the photon temperature
at matter-radiation equality.
This gives
\beq
T_d \simeq \left( \frac{D}{2} \right)^{2/3} \frac{(T_{reh})^2}{m_{pl}^{2/3} 
T_{eq}^{1/3}}
\eeq
The photon flux expected from graviton
decay is therefore
\beq
\frac{n_{G0}}{4 \pi} \frac{T_0}{T_d} \simeq 
 \left( \frac{2}{D} \right)^{2/3} \frac{N}{4 \pi} T_0^4 m_{pl}^{5/3}
T_{eq}^{1/3}
 \left( \frac{T_{reh}}{m_{pl (n+4)}} \right)^{n+2} (T_{reh})^{-3}
 \lappeq {\cal F} ~~~~.
\eeq
This gives
\beq
(T_{reh})^n < 3 \times 10^{-33} \frac{m_{pl (n+4)}^{n+2}}{{\rm GeV}^2}
~~~~~(60 ~{\rm MeV} < T_{reh} < 2 {\rm GeV})
\eeq
This applies for $\tau_U > \tau_G > t_{recomb}$, which
corresponds to the limit in parentheses (with $D=1$).

Photon number changing interactions
of the form $\gamma e \rightarrow e \gamma \gamma$ 
go out of equilibrium at $ t_{\gamma} \sim 10^{5}$ seconds.
If the gravitons decay after
$t_{\gamma}$, but before recombination, 
the photons they decay to will induce
a chemical potential\footnote{ The dimensionless parameter $\mu$ is defined as
the parameter in the Bose-Einstein distribution function: $ 1/ {(e^{\frac {E}{T} + \mu } +1)}$.} for the microwave
background \cite{DZ}:
\beq
\mu \simeq \frac{\rho_G}{\rho_{\gamma}}
\eeq
This  is in the instantaneous decay
approximation, where all the energy
of the gravitons is deposited into
the photons at $t = \tau_G$. This
should be a reasonable approximation for
$t_{\gamma} \ll \tau_G \ll t_{recomb}$
\cite{Salati}. The present experimental bound
is \cite{data} $\mu < 3.3 \times 10^{-4}$,
which implies
\beq
N \frac{m_{pl} T_{reh}^{n+2}}{m_{pl (n+4)}^{n+2}} \frac{T_{\gamma}^3}
{\rho_{\gamma}} < 3.3 \times 10^{-4}
\eeq
when the gravitons decay. The photon
temperature at decay $T_d$ can be determined
from $H(T_d)\sim \Gamma_G \simeq D T_{reh}^3/m_{pl}^2$,
so
one gets
\beq
m_{pl (n+4)}^{n+2} > 4 \times 10^{32} (T_{reh})^{n + 1/2} ~ {\rm GeV}^{3/2}
~~~~~(2 ~{\rm GeV} ~  \ll T_{reh} \ll 1 {\rm TeV})
\label{above}
\eeq
This applies for $10^{5} ~{\rm sec} \sim t_{\gamma} 
\ll \tau_G \ll t_{recomb} \sim 10^{13} $ seconds, or
$2 ~{\rm GeV} ~  \ll T_{reh} \ll 1$ TeV.

One of the successes of the Big Bang model is that
it predicts the correct abundances of light
elements. $^4He, ^3He, D$ and $^7Li$ are
synthesised in the early Universe
at temperatures just below an MeV, 
in about the right numbers to agree
with present observations \cite{BBN}.
If the gravitons decay  after
nucleosynthesis, one must check that the
their decay products do not 
destroy or produce  too many
of these light nuclei. This constraint
has been calculated for various particles
\cite{L,Ellis,DEHS}. There are  numerical bounds 
on $\rho_G / n_B$  in  \cite{DEHS} for
$ 10^4 ~{\rm sec} < \tau_G< 10^7 $ seconds,
 which  we can
simply translate into bounds on
$M_s$ as a function of $T_{reh}$.
These turn out to be similar or weaker 
than (\ref{above}).

\begin{figure}[htb]
\begin{center}
\epsfxsize=12cm \epsfysize=10cm \epsfbox{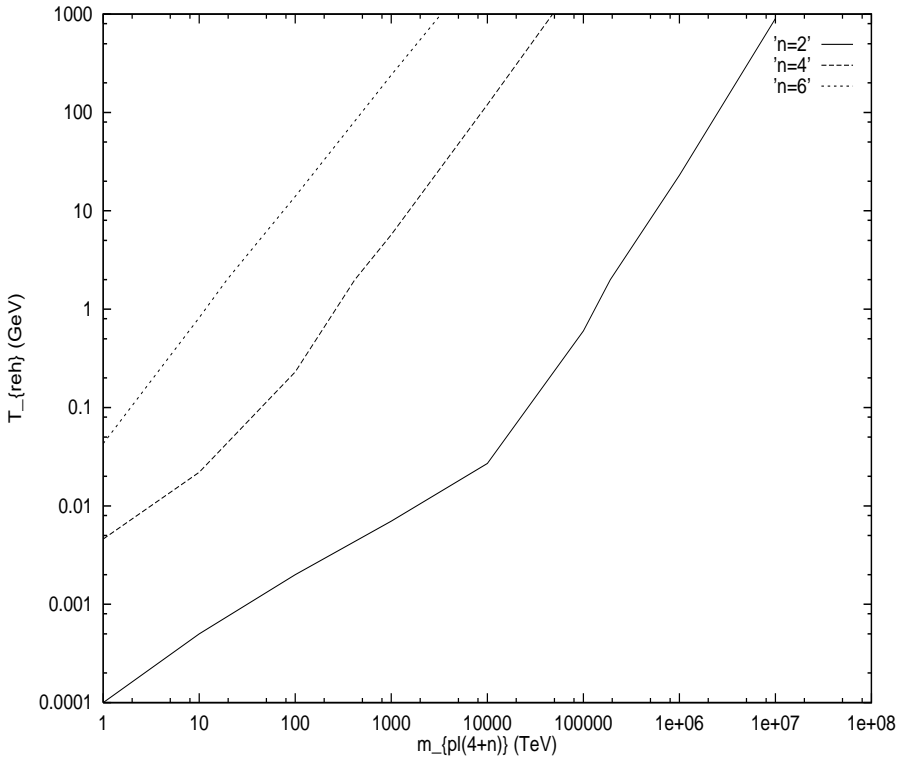}
\caption{Maximum allowed reheat temperature
$T_{reh} $ as a function of $m_{pl(n+4)}$
for different numbers $n$ of large extra dimensions.}
\end{center}
\label{f1}
\end{figure}

In figure 1 we plot the allowed reheat
temperature as a function of the $4+n$--dimensional Planck scale
$m_{pl (n+4)}$ for different numbers\footnote{ The case
of one extra dimension at the millimetre
leads to $T_{reh} \lappeq 10$ MeV,
which is easily compatible with primordial
nucleosynthesis.} of extra dimensions $n \geq 2$.
This is a fairly stringent bound; to
get a reheat temperature as large as
100 GeV, we need $m_{pl (n+4)}\sim 10^{6}$ GeV
for 6 extra large dimensions, and
$m_{pl (n+4)} \sim 10^{10}$ GeV for $n=2$.
If the reheat temperature is less than
100 GeV, electroweak baryogenesis \cite{EPTBAU}
and leptogenesis \cite{CoviB} (generating
a lepton asymmetry and then
having the ``sphalerons'' reprocess
it) are impossible.  If $T_{reh} \gg$ TeV, the
gravitons generated at $T_{reh}$ will
decay before nucleosynthesis and
thermalise rapidly, so they
are not a problem.

\section{Baryogenesis}

\subsection{Challenges for baryogenesis models}

First let us consider the consequences of the
low $T_{reh}$ constraint. For a large choice
of $M_s$ and of the number of large internal
dimensions, the reheat temperature must
be less than $\sim 100$ GeV, so the electroweak
$B+L$ violating processes are not available for baryogenesis. 
This means that electroweak
baryogenesis \cite{EPTBAU} and leptogenesis
\cite{CoviB} are not possible.
For larger values of $M_s$ and depending on $n$,  $T_{reh}\gappeq 100$ GeV is
 allowed and electroweak
baryogenesis is possible. This is attractive because the
non-perturbative electroweak $B+L$
violation proceeds through the operator
$(qqq\ell)^3$, which does not mediate
proton decay because it has $\Delta B = 3$
(as well as being exponentially small
at zero temperature).

There has recently been a very interesting suggestion \cite{Brand}
that the BAU could be generated at the QCD
phase transition using purely Standard Model
physics (the baryon number and CP violation
are spontaneous/non-perturbative). If this model works,
then one only needs a reheat temperature of order
1 GeV, which is easier to achieve than 100 GeV,
as one can see from figure 1. We do not
further discuss this mechanism, but it should be
kept in mind as a  possible way of
generating the baryon asymmetry
in low quantum gravity scale models.

The low $T_{reh}$  creates a
generic difficulty. One of the
Sakharov \cite{Sak} conditions for baryogensis
is that one needs some out-of-equilibrium
dynamics. This can be found at
phase transitions, or when
some interaction is not fast
enough to keep up with the expansion of the Universe.
However when the temperature
(or energy density) of the Universe is low,
the expansion rate is too ($H \sim
10^{-18}T$ at $T \sim $GeV), so
interactions have no difficulty
keeping up with the expansion.
Getting the out-of-equilibrium
anywhere but a phase transition is hard.
If the reheat temperature is less than 
$\sim 0.1 $ GeV, then the only phase transition
available appears to be the one out of inflation.

Another difficulty for baryogenesis models
is  the bounds on baryon number violation
discussed in section 2. For instance, to avoid 
fast proton decay through $|\Delta B| =|\Delta L| = 1$ operators,
and neutron-anti-neutron oscillations through $\Delta B = 2$ operators,
one may assume that $B$ is conserved
mod 3 . This is problematic  for scenarios
where the BAU is generated in the
out-of-equilibrium decay of a particle $X$.
 $X$ must have at least two
decay modes with different baryon number in the final
state, and approximately the same branching ratios \cite{KW}.
Otherwise the baryon asymmetry generated will
be small \footnote{This is a consequence of CPT: if
$X$  decays to a $B=B_1$ final state
with a large branching ratio $1 - \epsilon$, and a $B=B_2$
state with a small branching ratio $\epsilon$, then one can
assign $B=B_1$ to $X$, so the larger decay
is baryon number conserving. By CPT the total
decay rates of $X$ and $\bar{X}$ are equal, so
the baryon asymmetry created will be
proportional to $\epsilon - \bar{\epsilon}$
and therefore very small.}. 
If $B$ is conserved mod 3, then
$X$ must decay to final states with $ B = 1$ and
with  $ B = 2$ (or $B = 0$ and $B = 3$), so that
$X$ exchange generates   a vertex that
conserves $B$ mod 3. But  $B=2$ operators 
are of higher dimension that $B=1$ operators
(see table 1), so the branching ratio of $X$
to the $B=2$ final state will be very small. We tried
imposing $B$ mod 4,  so that $X$ could
decay via $\Delta B = 2$ and $\Delta B = -2$
processes, but  these operators are
of  dimension 10 and 12, so  that $X$ must have
a mass of order 100 GeV to decay before
nucleosynthesis...

If the quantum gravity scale is greater
than $10^5$ GeV in the SM
 ($10^9$ GeV in the MSSM)
then $\Delta B = 2$ operators do not
need to be suppressed/forbidden (see table 1). 
In this case, $B$ does not
need to be conserved, provided that
$L$ is; if there are only
baryon number violating couplings,
and the low energy theory
has Standard Model particle content,
 the proton cannot decay.
This means, for instance, that in SUSY models one can
use  the  interaction
$U^cD^cD^c$ to provide the baryon
number violation required for
baryogenesis. Such a  model 
of low reheat temperature baryogenesis
was constructed in \cite{DH}, where the inflaton
decay products include squarks, which then decay
via their $B$ violating coupling.
They decay before they have time
to thermalise or annihilate, so are out of equilibrium
and can generate a baryon asymmetry in their decay. 

\subsection{A contrived baryogenesis model}

Suppose that we are in the ``worst case scenario''
for baryogenesis.
This corresponds to the situation with $ M_s \lappeq 10^{5}$ GeV,
so  symmetries are required to forbid the
$n - \bar{n}$ operator $udsuds$,
and  the fast proton decay vertices.
The maximal allowed reheat temperature is much less than 100 GeV,
so there is no electroweak $B+L$ violation
available. If the motivation for having
a low $M_s$ is to solve the hierarchy problem,
we can also assume that there is no supersymmetry,
since this is also what it is for. This
means that Affleck-Dine baryogenesis is
not possible. Can
the baryon asymmetry be generated
in these circumstances?

We first try to construct an out-of-equilibrium
decay scenario. For this we need a particle
who decays out of equilibrium to final states
with different baryon numbers, with enough CP violation
in the decay rates to generate a baryon to
photon ratio $\eta \sim 3 \times 10^{-10}$.

Suppose $X$ is the inflaton. This has the advantage that
it decays
out of equilibrium. Moreover its width:
\beq
\Gamma_X \sim \frac{T_{reh}^2}{m_{pl}}
\label{22}
\eeq
must be small in order to obtain a low reheat temperature\footnote{We
assume that the inflaton is very weakly
coupled, so cannot decay by parametric resonance.}. 
One way to ensure that it has a long lifetime is to
make it decay via non-renormalisable operators. 
 For instance, this can happen via an operator of dimension
$4+d$ with coefficient $\lambda M_s^{-d}$, 
so that
$\Gamma \sim  \lambda^2 m_X^{2d+1}/M_s^{2d}$.
We   would like  $X$ to have baryon number
violating decays so that it can generate the baryon asymmetry,
which also means that $X$ should decay via non-renormalisable
interactions. 
As it oscillates about
its minimum, we suppose that
$m_X >$ GeV, so that it can produce  protons.

Another possibility is that $X$ is a particle 
generated in the reheating process, with
a number density $n_X = \delta n_{\gamma}$. 
The annihilation rate for $X$ will be
\beq
\Gamma_{ann} \sim n_X \sigma_{X\bar{X}\rightarrow anything}
\eeq
If we take $\sigma_{X\bar{X}\rightarrow anything} 
\sim  4 \pi\alpha^2/M_s^2$, then
requiring that $\Gamma_{ann} < H$ gives
\beq
4 \pi \alpha^2 \delta  <  \frac{M_s^2}
{T_{reh} m_{pl}}
\eeq
 If
we take   $M_s$ to
be   its minimum value $\gappeq 3 $ TeV
and $T_{reh}$ the maximum value
possible for $n \leq 6$ and $M_s < 10^5$ GeV
which is $ \lappeq 10$ GeV, then
this gives $\alpha^2 \delta < 10^{-14}$.
This is the condition such that   $X$ annihilations
will be out of equilibrium at the reheat temperature
and thereafter, so all the $X$s will decay.

We would like to address the possibility of having particles with
such small couplings. Consider for instance models obtained from Type
I'  strings after performing $T$-duality on all the internal
directions of a Type I model. There are two kind of $p$-branes in these models:
three-branes and seven-branes. We assume that the standard model lives
on the three-branes with gauge couplings of order one. The particles
that arise from seven-branes have gauge couplings suppressed by the
volume of the  four-dimensional internal space on which they are
wrapped. The corresponding couplings can be arranged to satisfy
the above constraints.

To generate the BAU,
$X$ needs similar branching
ratios to states with different
baryon number.
As discussed in the previous subsection,
this requirement is difficult to implement
in models where $B$ is conserved.
So instead we consider the possibility that
$B$ is not conserved, $L$ is conserved
mod 2 (which allows neutrino masses),
and there is a horizontal symmetry 
that suppresses the dangerous $\Delta B =2$
operators.

We assume that the SM
yukawa couplings are generated by
some horizontal $U(1)$ gauge symmetry \cite{hor},
which is spontaneously broken below 
$M_s$. The quarks ($q,u^c,d^c$) and leptons ($\ell,e^c$)
carry positive charges under this
symmetry, and the charges
are higher for the lighter
fermions.  The  Higgs
that breaks the horizontal
$U(1)$ with vev $\theta$ carries negative charge.
By choosing the horizontal charges of the fermions $Q^H_f$
with care, one can generate
approximately the right structure for the Yukawa
matrices, because the interaction
$u^c  u H \sim m_u \bar{u} u$ appears
multiplied by  $(\theta/\Lambda)^{Q^H_{u^c}+ Q^H_u}$
and $t^c t H \sim m_t \bar{t} t$ appears
multiplied by  $(\theta/\Lambda)^{Q^H_{t^c}+ Q^H_t}$. 
Such a mechanism is probably required
in models with a low $M_s$  to
avoid FCNC. It will also suppress
the problematic $u^cd^cs^cu^cd^cs^c$ operator:
at $M_s$ where $\theta$ is
zero, it is forbidden by the horizontal
symmetry (if all the fermions
are positively charged), and once
the horizontal symmetry is broken,
$u^cd^cs^cu^cd^cs^c$ can appear suppressed by
$(\theta/\Lambda)^{2(Q^H_{u^c} + Q^H_{d^c} +Q^H_{s^c})}$.
For $\theta/\Lambda \equiv \epsilon \sim .2$ and the
charges in table 2, the operator
$u^cd^cs^cu^cd^cs^c$ will be multiplied by
$\epsilon^{16}$, which 
is compatible with the experimental limit
for $\Lambda \gappeq $ few  TeV.
The proton is stable
enough provided that  $L$ is conserved mod 2.

\renewcommand{\arraystretch}{2}
\begin{table}
\begin{tabular}{||c||c|c|c|c|c||} 
\hline
\hline
generation &$ q$ &$u^c$ &$d^c$  & 
 $\ell$ & $e^c$ \\
\hline
1 & 5& 5& 2& 1& 5  \\ 
\hline
2 & 4& 2& 1& 0& 4 \\ 
\hline
3 & 1& 0& 1& 0& 1 \\
 \hline
 \hline
\end{tabular}
\label{t2}
\caption{Possible charges for the fermions and the Higgs
under the horizontal $U(1)$, for three generations.
The first generation is $u,d,e$, and so on. These charges generate
approximately the right Yukawa couplings.}
\end{table}

Suppose that $X$ is a light ($\sim 10$ GeV)
gauge singlet scalar
with $L=1$. It can decay to SM particles
via the dimension 7 operators $Xqqq\ell$ and $Xu^cu^cd^ce^c$.
These violate $B$ respectively
by 1 and -1 units, so a baryon asymmetry
could be generated.
We suppose that the fermions have the charges
under the horizontal $U(1)$ that are
listed in table 2. In this case
 the principle decay rates will be
\beq
\Gamma_{\bar{p}} \sim \epsilon^{18} \frac{m_{X}^7}{\Lambda^6}  ~~~~~~ 
 X \rightarrow c^c ~u^c ~b^c ~\tau^c ~~( \bar{D} ~\bar{B}~\bar{p} ~\tau^+) 
\eeq
\beq
\Gamma_p \sim \epsilon^{18}  \frac{m_{X}^7}{\Lambda^6}  ~~~~~~
 X \rightarrow c ~ s~ b~ \nu_{\tau} ~~ (D ~B~ K~ p ~\nu_{\tau}) 
\eeq
\beq
\Gamma_{p2} \sim \epsilon^{20}  \frac{m_{X}^7}{\Lambda^6}  ~~~~~~
 X \rightarrow c ~ d~b ~\nu_{\tau} ~~( D ~B~ p ~\nu_{\tau}) 
\eeq
where $\epsilon  = \theta/\Lambda \sim .2$.  We neglect 
kinematics, factors of $4 \pi$, and so on,
so these are very approximate estimates.
However, for $\Lambda \sim$ 3 TeV, and $T_{reh} \sim$
3 MeV,   equation
(\ref{22}) gives $m_X \sim 25$ GeV. This
is heavy enough to decay to $B$-- and $D$--mesons,
but light enough to (possibly) be produced in
the reheating process, or to be the inflaton. 
For larger $\Lambda$, we would need a larger $m_X$.

We have shown that we can construct
a scalar particle $X$ that  decays before 
nucleosynthesis, at about the
right time to reheat the Universe if it was the inflaton.
We now need to consider whether a sufficient
baryon asymmetry can be generated in the
decays. We assume that $\Gamma_p \gg \Gamma_{p2}$ so we neglect 
$\Gamma_{p2}$ and all the other smaller decay
modes. 
The net number of baryons produced per $X$ particle
will be
\beq
\frac{n_b}{n_X} \simeq  \frac{\Gamma_p - \Gamma_{\bar{p}}
+ \bar{\Gamma}_{\bar{p}} - \bar{\Gamma}_{p}}
{\Gamma_p + \Gamma_{\bar{p}}} 
 \equiv \theta_{CP}
\eeq
where
$\bar{\Gamma}$
is the CP conjugate decay.
The baryon-to-photon ratio $n_b/n_{\gamma}
\equiv \eta \simeq 3 \times
10^{-10}$ \cite{BBN} will be
\beq
\eta \simeq \frac{n_X}{n_{\gamma}}  \theta_{CP}~~~~.
\eeq
 If $X$ is
the inflaton, then $n_X/n_{\gamma} 
\sim T_{reh}/m_X \sim 10^{-3}$.
If $X$ is produced in the reheating
process, then ${n_X}/{n_{\gamma}} = \delta$
 is a model dependent parameter. One would
not expect to make more than one or two $X$s
in the decay of each inflaton, so in this case
 $\delta \lappeq  10^{-3}$.
This means that we need $\theta_{CP} \gappeq
10^{-7}$. If we assume that the CP violation
arises through loop corrections involving
new particles at  the scale $M_s$, then
$\theta_{CP} \sim (m_X/\Lambda)^2 \sim 10^{-6}$,
which is approximately right.

The family symmetry presented here  obviously suffers  from anomalies.
These might be cancelled in two different ways. The first is to assume
that massive particles in a hidden sector are 
charged under this $U(1)$, standard model 
symmetries and  some hidden gauge group. The hidden symmetry might 
suppress any undesirable non-renormalisable operator. Another possibility
is to appeal to a Green-Schwarz mechanism to cancel the anomaly \cite{hor}. 
If the 
gauge couplings are all given by the vacuum expectation value of a single
modulus (dilaton) then anomaly cancellation implies particular tree level
relations between the couplings. For the model at hand, 
the strong, weak and hypercharge $U(1)$ couplings are in the 
ratio $1:1:105/33$ at $M_s \sim$ TeV 
instead of the usual relation $1:1:5/3$ at $10^{16}$ GeV.  To compare
the tree level prediction with experimental 
measurements we need to know the precise evolution of coupling constants 
with energy from $M_s$ down to $M_Z$. Unfortunately
for low $M_s$ there is not yet a framework 
to discuss this
running of couplings as these become very sensitive to the spectrum
at energies of the order of TeV\footnote{See \cite{kar,DDG} 
for discussion of unification
in these models.}.

We also imposed $L$ as a 
spontaneously broken symmetry, to ensure 
proton longevity, so
 some additional (heavy)
leptons  must be included 
to cancel the anomalies
in $L$ \cite{IR}.

\subsection{Other possibilities}

It is clear from the previous section that
out of equilibrium decay scenarios do
not work easily at low scales with SM particle
content.
Electroweak baryogenesis and leptogenesis
will not work in their standard versions
if $T_{reh}$ is much
below the temperature at which
electroweak $B+L$ violation
is in equilibrium $\sim 100$ GeV.  However, there
are many other baryogenesis
mechanisms \cite{BAUrev}, some
of which may work naturally. We will
discuss these in a later publication
\cite{BD2}. The most popular mechanism
that we have not discussed is Affleck-Dine \cite{AD}.
This scenario is attractive for low string scale
models because the reheat temperature can generically
be low, and the dimension of the $B$ violating
operators is not so relevant. However,
the difficulty is that the vev should start with the
same phase over the whole observable Universe
\footnote{The CP violation in the Affleck-Dine
scenario is ``spontaneous'', that is encoded
in the relative phase between the vev
and the baryon number violating bumps
in the potential}. This is not so easy
to arrange if the expansion rate $H$
is much smaller than the flat
direction's mass $m \gappeq 100$ GeV,
because inflation cannot push the vev out
along the flat direction. It may
be possible to resolve this with
a small amount of external CP violation.
We will pursue this
possibility in a subsequent publication \cite{BD2}.

\section{Conclusion}

For traditional models, where the scale $M_s$ of quantum gravity lies
far away at energy scales of the order of $10^{19}$ GeV, 
the baryon asymmetry can
be generated in a plethora of scenarios. 
In contrast, we found that
 exhibiting
simple scenarios for  baryogenesis becomes a challenging problem when
$M_s \lappeq 10^5$ GeV. 
The three Sakharov requirements of
baryon number violation, $C$ and $CP$ violation,
and out-of-equilibrium must be satisfied.
Baryon number violation is hard to come by
because many baryon number violating
operators must be forbidden by a symmetry
to ensure that they are not  generated
at  $M_s$. Out-of-equilibrium
dynamics are also difficult because
there is
an  upper bound on
the reheat temperature  of the Universe
from  requiring that one not over-produce
gravitons in the extra large dimensions.
    We list experimental bounds on baryon number
violating operators in table 1, and  in figure 1 we plot
the maximum allowed reheat temperature as a function
of $m_{pl(4+n)}$ for different
numbers $n$ of large internal dimensions.
The $T_{reh}$ bound  could possibly be
avoided if the bulk fields (gravitons) could decay faster to hidden matter
whose energy redshifts.

  Standard electroweak baryogenesis  and leptogenesis
are excluded  for low $M_s$, because the reheat
temperature is constrained to be less than 100 GeV.
Affleck-Dine baryogenesis is difficult because
the Hubble expansion rate is not large enough
to drive the flat direction field out to a 
single  vev with the same phase everywhere. 

Out-of-equilibrium decay models are also  problematic;
the experimental bounds on baryon number violating
operators suggest  that baryogenesis
must proceed through  non-renormalisable operators of
very high dimension. 
An alternative is to  suppress baryon number violating
operators through  a horizontal family  symmetry,
and ensure that the proton remains stable
by  conserving $L$.  We
implement this idea in a toy 
model that could  generate 
the correct baryon asymmetry 
in the  
decay of a weakly coupled particle
(possibly the inflaton).

For larger values of $m_{pl(4+n)}$ we need
SUSY to solve the hierarchy problem,
in which case Affleck-Dine is a possibility.
If $m_{pl(4+n)} \gappeq  10^5$ GeV,
baryon number violation is allowed,
provided that $L$ is conserved.
For scales $M_s \gappeq 10^{10}$ GeV
the reheat temperature is large 
and electroweak baryogenesis is
possible.

We will return to discuss these issues in a future publication \cite{BD2}.

\subsection*{Acknowledgements}
S.D would like to thank Steve Abel for useful
conversations. The work of K.B. is supported by a John Bell scholarship
from the World Laboratory.


\begin{thebibliography}{222222}
\bibitem{nu98} for recent data and models
on the solar and atmospheric neutrino deficits, see {\it e.g.}
 {\it Neutrino 98},  Proceedings
of XVIII International Conference on Neutrino Physics
and Astrophysics, Takayama, Japan, 4-9 June 1998,
edited by Y. Suzuki, Y. Totsuka. (To appear as
{\it Nucl. Phys. B Proc. Suppl.}) 
\bibitem{BAUrev} for a review, see {e.g.}   A .D. Dolgov, 
  {\it  Phys. Rept.} {\bf 222} (1992) 309.
\bibitem{CMB} for a review, see {\it e.g.}
J. Silk, D. Scott and M. White,
{\it  Ann. Rev. Astron. Astrophys.} {\bf 32} (1994) 319.
\bibitem{TeVlim}  I. Antoniadis, {\it Phys. Lett.} {\bf B246} (1990) 377; I. Antoniadis and K. Benakli, {\it Phys. Lett.} {\bf B326} (1994) 69;
see also  V.A. Kostelecky and S. Samuel, {\it Phys. Lett.} {\bf B270} (1991) 21.
\bibitem{mmlim} E. Caceres, V. S. Kaplunovsky and I. Mandelberg,
 {\it Nucl. Phys.}{\bf 493}(1997) 73.

\bibitem{ADMR} I. Antoniadis, C. Mu\~noz and M. Quiros, {\it Nucl. Phys. } {\bf B397} (1993) 515;
 N. Arkani-Hamed, S. Dimopoulos and J. March-Russell, hep-th/9809124; R. Sundrum, hep-ph/9807348.

\bibitem{witten} E. Witten,  {\it Nucl. Phys. } {\bf B471} (1996) 135.

\bibitem{typeI} J. Lykken, {\it Phys. Rev. } {\bf D54} (1996) 3693;
I. Antoniadis, N. Arkani-Hamed, S. Dimopoulos and  G. Dvali, hep-ph/9804398.

\bibitem{ADD0} N. Arkani-Hamed, S. Dimopoulos and G. Dvali, 
{\it Phys. Lett.} {\bf B429} (1998) 263.

\bibitem{tye} G. Shiu and S.-H.H. Tye, hep-th/9805157.

\bibitem{kar} K. Benakli, hep-ph/9809582.

\bibitem{bachas0} C. Bachas, (1995) unpublished.
\bibitem{ABK} 
I. Antoniadis, K. Benakli and M. Quiros, {\it Phys. Lett.} {\bf B331} 
(1994) 313.
\bibitem{bachas} C. Bachas, hep-ph/9807415.


\bibitem{ADD} N. Arkani-Hamed, S. Dimopoulos and G. Dvali,
hep-ph/9807344.
\bibitem{APV} C.S. Wood {\it et al.}, {\it Science} {\bf 275} (1997) 1759. 

\bibitem{voisins} K.R. Dienes, E. Dudas, T. Gherghetta and A. Riotto,
hep-ph/9809406.
 
\bibitem{HW} P. Ho\v rava and E. Witten, {\it Nucl. Phys. } {\bf B460} (1996) 506; {\it Nucl. Phys. } {\bf B475} (1996) 94.
\bibitem{SUPERK} The Super-Kamiokande Collaboration, 
M. Shiozawa, {\it et al.,}
 KEK preprint 98-46, hep-ex/9806014;

\bibitem{proton} T. Banks and M. Dine,  {\it Nucl. Phys. } {\bf B479}
 (1996) 173; J. Ellis, A. Faraggi and D.V. Nanopoulos, {\it Phys. Lett.} 
{\bf B419} (1998) 123.

\bibitem{GR}  G.G.Ross, {\it Grand Unified Theories},
       Benjamin-Cummings, Menlo Park, CA.
\bibitem{PDG} {\it Review of Particle Physics},
C. Caso {\it et al.}, {\it Eur. Phys. J.} {\bf C3} (1998) 1. 
\bibitem{Zwirner}F. Zwirner,
{\it Phys.Lett.} {\bf 132B} (1983) 103. 
\bibitem{had}S.P. Misra and U. Sarkar, {\it Phys. Rev.}
{\bf D28} (1983) 249.
\bibitem{IR} L. E. Ibanez and G. G. Ross,
{\it Nucl.Phys.} {\bf B368} (1992) 3. 
\bibitem{hor} see, {\it e.g.}   M. Leurer, Y. Nir and N. Seiberg,
{\it  Nucl.Phys.} {\bf B398} (1993) 319-342;   L. E. Ibanez and G. G. Ross,
{\it Phys. Lett. } {\bf B332} (1994) 100; P. Binetruy and P. Ramond,
{\it Phys. Lett.} {\bf B350} (1995) 49.
\bibitem{T} M.S. Turner,
{\it  Boulder TASI 92}, 165;
 astro-ph/9304012.
\bibitem{RL}for a review, see {\it e.g.}
 D. Lyth and A. Riotto, to appear in
{\it Phys. Rep.} ; hep-ph/9807278 
\bibitem{lowT} P.Delbourgo-Salvador, P. Salati and J. Audouze,
{\it Phys.Lett.} {\bf B276} (1992) 115. 
\bibitem{BD2} S. Abel, K. Benakli and S. Davidson, in preparation.
\bibitem{Steve} S. Abel and S. Sarkar,
{\it Phys.Lett.} {\bf B342} (1995) 40, hep-ph/9409350.
\bibitem{BEN} K. Benakli, J. Ellis and D. Nanopoulos, hep-ph/9803333.
\bibitem{K+T}  E.R. Kolb and M. Turner {\it `` The Early Universe''},
Addison-Wesley, 1990, section 5.5.
\bibitem{DZ} for a review, see {\it e.g.}
 A.D. Dolgov and Ya.B. Zeldovich, {\it Rev. Mod. Phys.}
{\bf 53} (1981) 3; P.J.E. Peebles, {\it Physical Cosmology},
Princeton University Press, Princeton, USA.
\bibitem{Salati} P. Salati, {\it Phys. Lett.} {\bf B163} (1985) 236.
\bibitem{data} J.C. Mather {\it et al.,  Ap.J.} {\bf 420} (1994) 439.
\bibitem{BBN}  for a review, see {\it e.g.} G. Steigman,  
{\it  Nucl. Phys. Proc. Suppl.}, {\bf 48} (1996) 499;
  astro-ph/9602029.
\bibitem{L} D. Lindley, {\it Ap. J.} {\bf 294} (1985) 1.
\bibitem{Ellis} J. Ellis, D.V. Nanopoulos and S. Sarkar,
{\it Nucl. Phys.} {\bf B259} (1985) 175. 
\bibitem{DEHS} S. Dimopoulos, R Esmailzadeh, L.J.Hall and
G.D. Starkman, {\it Nucl. Phy.} {\bf B311} (1998) 699.
\bibitem{EPTBAU} for a review, see {\it e.g.}
  V.A. Rubakov and M.E. Shaposhnikov, 
  {\it  Usp. Fiz. Nauk} {\bf 166} (1996) 493-537 ( Phys. Usp. 39 (1996) 461); 
  hep-ph/9603208 
\bibitem{CoviB} for a review and references,
see {\it e.g.} W. Buchmuller and M. Plumacher,
In {\it Paris 1997, Phase transitions in cosmology} 141-160;  hep-ph/9711208. 
\bibitem{Brand} R. Brandenberger, I. Halperin and A. Zhitnitsky, hep-ph/980847.
\bibitem{Sak}  A.D. Sakharov,   {\it JETP Lett.} {\bf 5} (1967) 24. 
\bibitem{KW} E. Kolb and S. Wolfram,
{\it Nucl. Phys.} {\bf B172} (1980) 224; Erratum-{\it ibid.} {\bf B195}
(1982) 542. 
\bibitem{DH} S. Dimopoulos and L. Hall, {\it Phys. Lett.}
{\bf B196} (1987) 135. 

\bibitem{DDG} K. R. Dienes, E. Dudas and T. Gherghetta, {\it Phys. Lett.}
{\bf B436} (1998) 55; hep-ph/9806292.

\bibitem{AD} see, for instance, 
  I. Affleck and  M.Dine, {Nucl. Phys.} {\bf B 249}
  (1985) 361 ; 
  J. Ellis , K. Enqvist, D.V. Nanopoulos and K.A. Olive, 
  {\it Phys. Lett.} {\bf  B191} (1987) 343;
  M.Dine, L. Randall and S. Thomas, {\it Phys. Rev. Lett.}
  {\bf 75} (1995) 398;
  {\it ibid.} {\it Nucl. Phys.} {\bf B458} (1996) 291;
  M. Gaillard, H. Murayama and K.A. Olive, {\it Phys.Lett.} {\bf B355}
  (1995) 71.
\end{thebibliography}
\end{document}